\documentclass[handcarry]{aiaa-tc}
\usepackage{tikz}
\usetikzlibrary{calc}
\usepackage{amssymb}
\usepackage{amsmath}
\usepackage{url}
\DeclareMathOperator{\sign}{sign}

\graphicspath{{./figures/}}

\newcommand{\dario}{\xi}

 \title{On the astrodynamics applications of Weierstrass elliptic and related functions}

 \author{
  Dario Izzo%
    \thanks{Scientific Coordinator, Advanced Concepts Team, European Space Agency, Noordwijk, The Netherlands.}\\
  {\normalsize\itshape
  European Space Agency, Noordwijk, 2201AZ, The Netherlands}\\
  \and
  Francesco Biscani%
   \thanks{European Space Agency, Noordwijk, 2201AZ, The Netherlands}\\
  {\normalsize\itshape European Space Agency, Noordwijk, 2201AZ, The Netherlands}
 }

 \AIAApapernumber{AAS 16-279}
 \AIAAconference{AAS/AIAA Space Flight Mechanics Meeting, Napa, CA in February 14-18, 2016}

\begin{document}

\maketitle

\begin{abstract}
Weierstrass elliptic and related functions have been recently shown to enable analytical explicit solutions to classical problems in astrodynamics. These include the constant radial acceleration problem, the Stark problem and the two-fixed center (or Euler's) problem. In this paper we review the basic technique that allows for these results and we discuss the limits and merits of the approach. Applications to interplanetary trajectory design are then discussed including low-thrust planetary fly-bys and the motion of an artificial satellite under the influence of an oblate primary including $J_2$ and $J_3$ harmonics. 
\end{abstract}

\section*{Nomenclature}

\begin{tabbing}
  XXXXXXXX \= \kill
  $f$ \> Generic function, polynomial \\
  $P$ \> Generic 3rd or 4th order polynomial \\
  $z, w$ \> Complex variables \\
  $\Im$ \> Imaginary part of a complex number \\
  $\Re$ \> Real part of a complex number \\
  $\wp$ \> Weierstrass elliptic function \\
  $\wp^{-1}$ \> Inverse of the Weierstrass elliptic function \\
  $\zeta$ \> Weierstrass Zeta function \\
  $\sigma$ \> Weierstrass Sigma function \\
  $g_2, g_3$ \> Lattice invariants \\
  $\omega_1, \omega_2, \omega_3$ \> Half-periods of the Weierstrass elliptic and related functions \\
  $e_1, e_2, e_3$ \> Lattice roots \\
  $\tilde e_1, \tilde e_2, \tilde e_3 (, \tilde e_4)$ \> Roots of the third (or fourth) order polynomial\\
  $\tau$ \> Sundmann transformed pseudo-time (or anomaly) \\
  $\tau_m$ \> Pseudo-time of root passage \\  
  $A, B$ \> Constants \\  
  $\mathcal E$ \> Specific energy \\
  $\mathcal T$ \> Kinetic energy \\
  $\mathcal V$ \> Potential \\
  $h$ \> angular momentum \\
  $a$ \> semi-major axis \\
  $e$ \> eccentricity \\
  $p$ \> orbital parameter \\
  $\gamma$ \> flight path angle \\
  $\delta$ \> asymptote deflection angle \\
  $\mu$ \> gravitational parameter \\
  $\alpha$ \> spacecraft acceleration \\
  $r, v$ \> radius and velocity magnitudes \\
  $\xi, \eta$ \> parabolic coordinates \\
  $\theta$ \> true anomaly \\[5pt]

  \textit{Special notation}\\
  $\mathbf r, \mathbf v$ \> vectors \\
  $f'$ \> derivative \\
  $\dot f$ \> time derivative \\[5pt]
  
  \textit{Subscripts}\\
  $m$ \> referred to a polynomial root (root passage, pericenter) \\
  $K$ \> Keplerian \\
  $\infty$ \> At infinity
 \end{tabbing}

\section{Introduction}
Weierstrass elliptic and related functions appear in the solution of many problems in physics. In general relativity, for example, they are an established tool to tackle complex issues \cite{biscani_first-order_2013, gibbons_application_2012, scharf_schwarzschild_2011, hackmann_analytical_2010}. In astrodynamics, only recently, they have been used to find explicit solutions to three fundamental problems: the constant radial acceleration problem\cite{radial}, the Stark problem\cite{stark} and the two fixed center problem\cite{euler}, also known as Euler's three body problem. The constant radial acceleration problem consists in describing the motion of a point mass particle subject to a central gravity field and to an additional constant radial acceleration. The Stark problem consists in describing the motion of a point mass particle subject to a central gravity field and to an additional acceleration constant in the inertial reference frame. The Euler's three body problem consists in describing the motion of a point mass particle subject to the gravity field of two masses fixed in the inertial frame. In all cases, the resulting dynamical system is integrable in the Liouville sense and in all cases the resulting dynamics, extensively studied both from a theoretical and an applicative perspective, admits an explicit analytical solution via the use of Weierstrass functions and the introduction of anomalies (or pseudo-times) introduced via Sundmann transformations. The actual solution in the real time is recovered by solving a transcendental function of such anomalies or pseudo-times (analogues to Kepler's equation). It is interesting to both note and further study the close analogy to the solution to Kepler's problem. Indeed the procedures and expressions are, at least formally, analogues of the Keplerian ones: the main difference being in their use of Weierstrass elliptic and related functions rather than of circular functions. 

In this paper we review the generic solution procedure that allowed to obtain these results and we discuss the computer implementation of the new resulting procedures. We start with a basic introduction to Weierstrass elliptic and related functions and their relation to applications in astrodynamics. We then discuss the computer implementation of these functions showing how their evaluation cost is, essentially, comparable to that of circular functions when we restrict the evaluation to the real axis and assume the lattice properties as known. In the following section the constant radial acceleration problem is considered and a procedure to solve the initial value problem is detailed where evaluations of the Weierstrass functions are kept in the real domain in most of the cases. Then, the case of a radially powered planetary fly-by is studied in detail and an expression returning the asymptote deflection angle is developed. In the following section we turn our attention to the 2D Stark problem, first deriving the full solution, and then studying a second case of powered fly-by, here called the Stark fly-by. We find analytical expression that allow to design such a fly-by with ease thus allowing to study the exploitation of Oberth effect in a low-thrust trajectory. In the final section, we briefly show how the problem of artificial satellite motion under a gravity field including the $J_2$ and $J_3$ perturbations also has an analytical, explicit solution in terms of the Weierstrass functions.

\section{Weierstrass elliptic and related functions}
Weierstrass elliptic and related functions are a group of special functions that were studied and introduced by Karl Weierstrass at the end of the 19th century as an improvement over the Jacobian elliptic functions $\mbox{sn}$, $\mbox{cn}$ and $\mbox{dn}$. Today it is accepted \cite{bianchi} that they constitute a superior tool to construct a generic theory of elliptic functions and that they are often advantageous to solve integrals in the form:
$$
\int f\left(x, \sqrt{P(x)}\right) dx
$$
where $f$ is a function of $x$ and of the square root of a third or fourth order polynomial $P$ not resolved into factors (Byrd\cite{byrd} pag. 1889) (if the polynomial is resolved into known fixed factors, the Jacobian approach often offers a valid alternative). We report briefly the definition of these functions and a few theorems that establish their fundamental relation to fundamentals problems in astrodynamics. We follow and use the conventions and developments discussed extensively in the on-line version of the NIST Handbook of Mathematical Functions\cite{dlmf}.

\subsection{Definition}
\begin{figure}[t!]
  \centering
  \begin{tikzpicture}[scale=0.6]
    \coordinate (Origin)   at (0,0);
    \coordinate (XAxisMin) at (-3,0);
    \coordinate (XAxisMax) at (5,0);
    \coordinate (YAxisMin) at (0,-2);
    \coordinate (YAxisMax) at (0,5);
    \draw [thin, gray,-latex] (XAxisMin) -- (XAxisMax);
    \draw [thin, gray,-latex] (YAxisMin) -- (YAxisMax);

    \clip (-5,-5) rectangle (4cm,4cm); 
    \pgftransformcm{1}{0.0}{0.0}{1.5}{\pgfpoint{0cm}{0cm}}
    \coordinate (Bone) at (0,1);
    \coordinate (Btwo) at (1,-1);
    \draw[style=help lines,dashed] (-14,-14) grid[step=2cm] (14,14);
    \foreach \x in {-7,-6,...,7}{
      \foreach \y in {-7,-6,...,7}{
        \node[draw=gray,circle,inner sep=1pt,fill=gray] at (2*\x,2*\y) {};
      }
    }
    \draw [thick,-latex,red] (Origin)
        -- (Bone) node [above left] {$\omega_3$};
    \draw [thick,-latex,red] (Origin)
        -- ($(Bone)+(Btwo)$) node [below right] {$\omega_1$};
    \draw [thick,-latex,red] (Origin)
        -- (-1,-1) node [left] {$\omega_2$};
    \filldraw[fill=gray, fill opacity=0.3, draw=black] (Origin)
        rectangle ($2*(Bone)+(Btwo)$);
    \filldraw[fill=gray, fill opacity=0.3, draw=black] (Origin)
        rectangle ($4*(Bone)+2*(Btwo)$);
        
    \node[] at (1,-2.5) {$\Delta > 0$};

  \end{tikzpicture}
  \hspace{1cm}
  \begin{tikzpicture}[scale=0.6]
    \coordinate (Origin)   at (0,0);
    \coordinate (XAxisMin) at (-3,0);
    \coordinate (XAxisMax) at (5,0);
    \coordinate (YAxisMin) at (0,-2);
    \coordinate (YAxisMax) at (0,5);
    \draw [thin, gray,-latex] (XAxisMin) -- (XAxisMax);
    \draw [thin, gray,-latex] (YAxisMin) -- (YAxisMax);

    \clip (-5,-5) rectangle (4cm,4cm); 
    \pgftransformcm{1}{-1}{1}{1}{\pgfpoint{0cm}{0cm}}
    \coordinate (Bone) at (0,1);
    \coordinate (Btwo) at (1,-1);
    \draw[style=help lines,dashed] (-14,-14) grid[step=2cm] (14,14);
    \foreach \x in {-7,-6,...,7}{
      \foreach \y in {-7,-6,...,7}{
        \node[draw=gray,circle,inner sep=1pt,fill=gray] at (2*\x,2*\y) {};
      }
    }
    \draw [thick,-latex,red] (Origin)
        -- (Bone) node [above left] {$\omega_3$};
    \draw [thick,-latex,red] (Origin)
        -- (-1,-2) node [below right] {$\omega_2$};
    \draw [thick,-latex,red] (Origin)
        -- (1,1) node [midway, below] {$\omega_1$};
    \filldraw[fill=gray, fill opacity=0.3, draw=black] (Origin)
        rectangle ($2*(Bone)+(Btwo)$);
    \filldraw[fill=gray, fill opacity=0.3, draw=black] (Origin)
        rectangle ($4*(Bone)+2*(Btwo)$);
        
    \node[] at (1,-3) {$\Delta < 0$};
  
  \end{tikzpicture}
  \caption{The two possible lattices in $\mathbb{C}$ having real invariants $g_2$ and $g_3$ }
  \label{fig:lattices}
\end{figure}
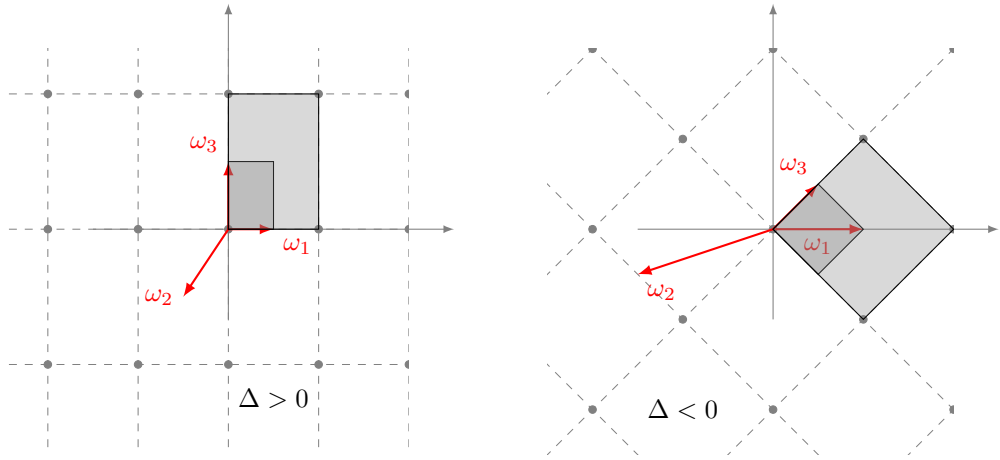
Consider any pair of complex numbers $\omega_1, \omega_3 \in \mathbb C$ such that $\Im(\omega_3 / \omega_1) > 0$. This last request makes sure that one can rotate counterclockwise $\omega_1$ until an overlap to $\omega_3$ spanning less than $180$ degrees. The set of points $2n\omega_1 + 2m\omega_3$ with $n,m\in \mathbb{Z}$ define a lattice $\mathbb{L}$ in the complex plane. The quantities $2\omega_1$ and $2\omega_3$ are called lattice generators and are not unique. If, for example,  $\omega_1 + \omega_2 + \omega_3 = 0$, then $2\omega_3, 2\omega_2$ and $2\omega_2, 2\omega_1$ are also generator of the same lattice $\mathbb{L}$. Weierstrass defined the following function:
$$
\wp(z|\mathbb L) = \frac{1}{z^2} + \sum_{w\in\mathbb L \backslash \left\{0\right\}}\left( \frac{1}{(z-w)^2} - \frac{1}{w^2} \right)
$$
where the series is uniformly and absolutely convergent, and thus the exact order of its terms is irrelevant. When not needed, the underlying lattice $\mathbb L$ is omitted from the notation. From the above definition it follows that $\wp(z+2\omega_i) = \wp(z)$, hence $\wp$ is a doubly periodic function in $\mathbb{C}$, that is $\wp$ is an elliptic function. In a similar way, Weierstrass introduced two more functions, $\sigma(z|\mathbb{L})$ and $\zeta(z|\mathbb{L})$, quasi-periodic, thus not elliptic, and having the following differential relations to $\wp$:
$$
\begin{array}{l}
    \wp(z) = \zeta'(z) \\
    \zeta(z) = \sigma'(z) / \sigma(z)
\end{array}
$$
The following quantities are the so-called lattice invariants:
$$
\begin{array}{l}
g_2 = 60 \sum_{w\in\mathbb{L}\backslash \left\{0\right\}} w^{-4}\\
g_3 = 140 \sum_{w\in\mathbb{L}\backslash \left\{0\right\}} w^{-6}
\end{array}
$$
they are constants defined as a sum over all the lattice points except the origin and are thus determined solely by the lattice itself. Each couple of lattice generators (for example $\omega_1$, $\omega_3$) thus determine univocally the lattice invariants. Conversely, given a couple of complex numbers, there is only one lattice $\mathbb{L}$ having them as invariants. The polynomial $g(w) = 4w^3 - g_2 w -g_3$, factorized as $g(w) = (w-e_1)(w-e_2)(w-e_3)$, defines the lattice roots $e_i$. Given any pair of generators $w_1, w_3$ and $w_2 = -w_1-w_3$, the lattice roots can be ordered and identified through the relation:
$$
\wp(\omega_i) = e_i
$$
which we will assume valid in all of the following developments. In most of the applications that we are interested in, the lattice invariants are real numbers. From the lattice invariants definition it is immediate to see that when a lattice is symmetric with respect to the real axis then necessarily $g_2, g_3 \in \mathbb{R}$. It is possible to show that this is also a sufficient condition so that only two types of lattices will be possible resulting in real lattice invariants and are visualized in Figure \ref{fig:lattices}. The following relations derive from the identity $4w^3 - g_2 w -g_3 = (w-e_1)(w-e_2)(w-e_3)$ and link the lattice roots to its invariants :
$$
\begin{array}{l}
e_1+e_2+e_3 = 0 \\
g_2 = -4(e_1e_2+e_1e_3+e_2e_3) \\
g_3 = 4e_1e_2e_3
\end{array}
$$
The discriminant $\Delta = g_2^3-27g_3^2$ determines whether the roots will be all real and distinct ($\Delta>0$) or one real and two complex conjugates ($\Delta < 0$), as well as the lattice type (see Figure \ref{fig:lattices}). Note that under the selected convention $\omega_1$ is always real and $\omega_3$ is either a pure imaginary number ($\Delta > 0$, the lattice roots are all real) or a complex quantity with positive imaginary part ($\Delta < 0$, only one lattice root is real). The fundamental result revealing the importance of lattice invariants is the differential identity:
$$
\wp'^2 = 4\wp^3 - g_2\wp-g_3
$$
which also implies the important integral definition for $\wp$,
$$
z = \int_{\wp(z)}^\infty (4s^3 - g_2s - g_3)^{- \frac 12}ds
$$

\begin{figure}[t!]
\centering
\includegraphics[width=0.95\columnwidth]{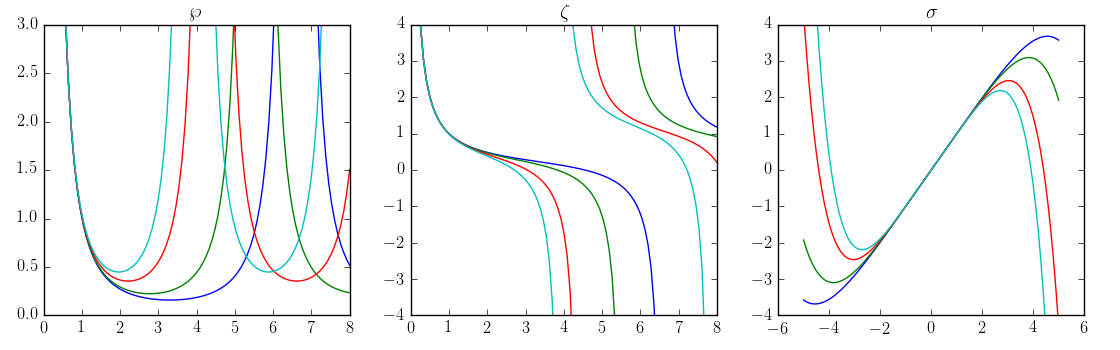}
\caption{Plot of the Weierstrass functions $\wp(x), \zeta(x)$ and $\sigma(x)$ for $g_3=0$ (lemniscatic case), $g_2=[0.1,0.2,0.5,0.8]$ and a real $x$. \label{fig:functions}}
\end{figure}

\subsection{Relevance to Astrodynamics}
The importance of Weierstrass elliptic and related functions to astrodynamics can be best appreciated considering the following integral definition of a function $\tau$:

\begin{equation}
\tau(x) - \tau_m = \pm \int_{x_m}^{x} \frac{1}{\sqrt{a_0s^4 + 4a_1s^3 + 6a_2 s^2 + 4a_3 s + a_4}} ds
\label{eq:weierstrass1}
\end{equation}
which appears in many fundamentals problems of astrodynamics as the relation between a pseudo-time $\tau$ (or an anomaly) and a state variable $x$. A generic procedure to solve the above integral and find $x(\tau)$ rather than $\tau(x)$ is well described in the classic book from Whittaker and Watson\cite{whittaker} (see \S 20.6), and results, when $a_0$ and $a_1$ are not both null and the polynomial $f = a_0s^4 + 4a_1s^3 + 6a_2 s^2 + 4a_3 s + a_4$ has no repeated factors, in the use of the following inversion formula:
\begin{equation}
x(\tau) = x_m + \frac{A}{\wp(\tau - \tau_m, g_2, g_3)-B}
\label{eq:weierstrass2}
\end{equation}
where $x_m$ is a root of $f$, $A = 1/4 f'(x_0)$,  $B = 1 / 24 f''(x_0)$ and the two lattice invariants are:
\begin{equation}
\begin{array}{l}
g_2 = a_0a_4 - 4a_1a_3+3a_2^2 \\
g_3 = a_0a_2a_4+2a_1a_2a_3-a_2^3-a_0a_3^2-a_1^2a_4
\end{array}
\label{eq:lattice_invariants}
\end{equation}
The quantity $\tau_m$ is what we call \lq\lq time of root passage\rq\rq\ since $x(\tau_m) = x_m$, and can be computed writing Eq.(\ref{eq:weierstrass2}) at $\tau=0$ and finding $\wp$ from it:
\begin{equation}
\wp(\tau_m) = B + \frac{A}{x_0 - x_m}
\label{eq:timeofrootpassage}
\end{equation}
where $x_0$ is the initial value $x(0)$. The inversion of the Weierstrass function will return two valid values for $\tau_m$ which reflect the original ambiguity in the integral sign. Such an ambiguity is solved forcing the initial condition. The derivative of Eq.(\ref{eq:weierstrass2}) with respect to $\tau$ is:
\begin{equation}
x'(\tau) = -\frac{(x(\tau) - x_m)^2}{A}\wp'(\tau - \tau_m)
\label{eq:weierstrass3}
\end{equation}
which, for $\tau=0$, holds:  
$$
\wp'(\tau_m) = \frac{Ax'_0}{(x_0-x_m)^2}
$$ 
which can be used to univocally determine $\tau_m$ from the initial condition $x_0'$. 
It is noteworthy that any root $x_m$ of the polynomial $f(s)$ can be used with the above formulae be it a real, pure imaginary or complex root. Choosing a real root, when possible, has two main advantages: it allows to define the origin of the $\tau$ pseudo-time variable so that $\tau_m=0$, and it keeps all the computations in the real domain (i.e. no complex quantities involved) resulting in a significant increase in the efficiency of evaluating the expressions on a computer as shown in the next section.

\subsection{Notes on the computer implementation}
\label{sec:computational efficiency}
\begin{table}[t!]
\begin{center}
\begin{tabular}{l||c|c|c}
Function & type & evaluations & ns / evaluation \\
\hline
real $\sin$& double&                    20000000&          62 \\
real $\wp$& double&                     10000000 &        178 \\
real $\wp'$& double&                    5000000 &        215 \\
real $\zeta$& double&                   10000000&         166 \\
real $\sigma$& double&                  10000000 &        175 \\
complex $\sin$& double&                 10000000 &        175 \\
complex $\wp$& double&                  2000000 &       532 \\
complex $\wp'$& double &                2000000 &       787 \\
complex $\wp^{-1}$& double&             1000000  &      1643 \\
complex $\zeta$& double&                2000000  &      531 \\
complex $\sigma$& double &              2000000  &      659 
\end{tabular}
\end{center}
\caption{Performances of Weierstrass elliptic and related functions as implemented in the project w\_elliptic. Results for the standard sin function (std::sin) are also shown for comparison.\label{tab:perf}. The table has been obtained running the project test suite compiled using gcc using the -ffast-math flag on a Intel(R) Core(TM) i7-3610QM CPU.}
\end{table}

The computer implementation of Weierstrass elliptic and related functions received little or no attention from the computer science community in the last decades. As a consequence,  not many languages nor tools are offering the possibility to compute these functions and it is very difficult to assess their use in terms of computational efficiency, for example with respect to the much more studied and popular Jacobi's elliptic functions. An attempt was recently made, limited to the Stark problem, by Hatten and Russel\cite{hatten} who, not having access to efficient implementations of the Weierstrass functions nor of the expressions using them, were forced to conclude that Weierstrass functions carry a computational penalty with respect to other methods. Such a conclusion was also later used to justify the methodology adopted by Beth et al. in their study on planetary exospheres\cite{beth1, beth2}. In reality, while it is true that the construction of the lattice from the invariants $g_2, g_3$ is a necessary step whose cost must be paid, once this step is performed the actual evaluation of the Weierstrass elliptic and related function is extremely fast as shown in Table \ref{tab:perf} where their speed is compared to the computational speed of the simplest trigonometric function (std::sin). In order to build such a table we programmed a C++ (and python) open source project called w\_elliptic (\url{https://github.com/bluescarni/w_elliptic}) that implements efficient versions of these functions distinguishing with respect to the argument type being a complex or a real (double) number. It is not in the scope of this paper to discuss the implementation details of w\_elliptic, which is still undergoing further optimization and improvements and is the subject of a dedicated paper under preparation\cite{inprep}. It is, though, very clear that these functions can be computed extremely efficiently, especially in the real domain. We will see in the rest of this paper how indeed most of the expressions involved in the solution of fundamental astrodynamical problems can be written as to keep the argument of the Weierstrass functions in the real domain. Any use of explicit solutions in terms of Weierstrass functions using their complex implementation is bound to be much slower as shown in the Table. 

\section{Solution to the constant radial acceleration initial value problem}
Previous work \cite{radial} reported the fundamental theoretical developments that lead to solve explicitly the radial acceleration problem. We here use those developments to establish a procedure to find the spacecraft position and velocity  $\mathbf r, \mathbf v$ at any time $t$.

Define the motion invariants, i.e. the angular momentum and the specific energy:
\begin{equation}
h = r_0^2 \dot\theta = \mathbf r_0 \times \mathbf v_0
\label{eq:cons_h}
\end{equation}
\begin{equation}
\mathcal E = \frac{v_0^2}2 - \frac \mu{r_0} - \alpha r_0
\label{eq:vis_viva}
\end{equation}
Note that $\alpha > 0$ corresponds to an outward pointing acceleration.
Compute the two invariants of the Weierstrass elliptic and related functions in the constant radial acceleration case:
\begin{equation}
\begin{array}{l}
g_2 = \frac{\mathcal E^2}{3} - \alpha \mu \\
g_3 = \frac{\alpha^2}{4}(h^2 + \frac{2 \mathcal E \mu}{3\alpha} - \frac{4\mathcal E^3}{27\alpha^2})
\end{array}
\end{equation}
Define the following third order polynomial and its derivatives:
$$
f(r) = 2\alpha r^3 +2 \mathcal E r^2 + 2 \mu r - h^2
$$
$$
f'(r) = 6\alpha r^2 +4 \mathcal E r + 2\mu
$$
$$
f''(r) = 12\alpha r +4 \mathcal E
$$
Compute the three roots $\tilde e_1, \tilde e_2, \tilde e_3$ of $f$. Compute the radius of pericenter passage $r_m$ as the real root closest to $r_0$ and such that $r_m \le r_0$. Using the energy conservation equation define the velocity of pericenter passage $v_m^2 = 2(\mathcal E + \frac \mu {r_m} + \alpha r_m)$. Compute the constants $B = \frac 1{24} f''(r_m)$, $A = \frac 14 f'(r_m)$ and the quantity $\xi$ defined as $\wp(\xi) = (B - \frac{A}{r_m})$, $\wp'(\xi) = \frac{Ah}{r_m^2} \imath $. Note that $\xi$ will be a real number if $A < 0$.

Introducing the radial anomaly $\tau(t)$ (via a Sundmann transformation $\frac{dt}{d\tau} = r$), the final explicit solution is given by:
\begin{equation}
\label{eq:radial1}
\left\{
\begin{array}{l}
r(\tau) = r_m + \frac{A}{\wp(\tau) - B} \\
\sin \theta(\tau) = x(\tau) \sin(v_m\tau) - y(\tau) \cos(v_m \tau) \\
\cos \theta(\tau) = y(\tau ) \sin(v_m\tau) + x(\tau) \cos(v_m \tau)
\end{array}
\right.
\end{equation}
where we defined $x(\tau) + iy(\tau) = \frac{\sigma(\dario-\tau)}{\sigma(\tau+\dario)} \exp{2\tau\zeta(\dario)}$. We also have:
\begin{equation}
\label{eq:radial2}
\left\{
\begin{array}{l}
\dot r(\tau) = - \frac{(r(\tau) - r_m)^2}{Ar(\tau)} \wp'(\tau) \\
\dot \theta(\tau) = \frac{h}{r^2(\tau)}
\end{array}
\right.
\end{equation}
The relation between the time and the radial anomaly is described via the radial Kepler's equation which admits the following equivalent forms:
\begin{equation}
t(\tau) = r_m \tau  - \frac{ 4 B A}{g_3+ 8 B^3}\left[ B\tau +
\zeta(\tau) + \frac 12 \frac{\wp'(\tau)}{\wp(\tau) - B} \right]
\label{eq:radial_kepler}
\end{equation}
\begin{equation}
t(\tau) = r_m \tau  - \frac{ 2 B A}{g_3+ 8 B^3}\left[ 2 B\tau +
\zeta(\tau-\omega_k) + \zeta(\tau + \omega_k) \right]
\label{eq:radial_kepler1}
\end{equation}
where $\wp(\omega_k) = B$. Note that in case the motion is bounded, necessarily $\wp(\tau) > B, \forall \tau \in R$, hence $\omega_k$ will be a complex quantity, while in case of unbounded motion $\omega_k$ will necessarily be a real quantity and thus the second expression, as the first one, would also involves only real quantities while is not undefined at $\tau = 0$. Note that both radial and true anomaly $\tau$, $\theta$ are zero at a pericenter passage, i.e. $t=0 \rightarrow \tau=0$, $\theta = 0$, $r = r_m$ . The radial anomaly at the initial conditions is found computing the expression:
\begin{equation}
\wp(\tau_0) = B - \frac{A}{r_m-r_0}
\label{eq:tau0}
\end{equation}
while the true anomaly at the initial conditions is found computing the expression:
\begin{equation}
\begin{array}{l}
\sin \theta_0 = x(\tau_0) \sin(v_m\tau_0) - y(\tau_0) \cos(v_m \tau_0) \\
\cos \theta_0 = y(\tau_0) \sin(v_m\tau_0) + x(\tau_0) \cos(v_m \tau_0)
\end{array}
\label{eq:theta0}
\end{equation}
The explicit Cartesian coordinates of the satellite can be finally obtained, trivially, using the following expression:
\begin{equation}
\begin{array}{l}
\mathbf r(t) = r(t)\cos\theta(t) \mathbf i_{r_m} + r(t)\sin\theta(t)\mathbf i_{p_m} \\
\mathbf v(t) = (\dot r(t)\cos\theta(t) - \frac h{r(t)} \sin\theta(t)) \mathbf i_{r_m} + (\dot r(t)\sin\theta(t) +\frac h{r(t)} \cos\theta(t))\mathbf i_{p_m}
\end{array}
\end{equation}
which involves the solution to the Kepler's radial equation in order to get $t$ from the radial anomaly $\tau$.

\section{Radial fly-by}
\begin{figure}[t!]
\begin{center}
\begin{tikzpicture}[scale=5]
    \pgfmathsetmacro{\e}{1.1}   
    \pgfmathsetmacro{\a}{1}
    \pgfmathsetmacro{\b}{(\a*sqrt((\e)^2-1)} 
    
    \def\offset{0.00} 
    
    \path [draw=none,fill=gray, fill opacity = 0.1] (-\e*\a,0) circle (0.7);
    \path [draw=none,fill=white] (-\e*\a,0) circle (0.4);    
    
    \draw[->] (-\e*\a-\offset,0,0) -- (0.25-\e*\a-\offset,0,0) node[right] {$\mathbf i_{r_m}$};
    \draw[->] (-\e*\a-\offset,0,0) -- (-\e*\a-\offset,0.25,0) node[left] {$\mathbf i_{p_m}$};
    
    \draw[] (-\e*\a-\offset,0) -- ({-\a*cosh(-1)},{\b*sinh(-1)}) node[midway, above, fill=white] {$r_o$};
    
    \draw[] (-\e*\a-\offset,0) -- ({-\a*cosh(-0.725)},{\b*sinh(-0.725)}) node[midway, right, fill=white] {$r_i$};
        
    \draw[arrows=<->](-\e*\a-\offset,-0.02)--(0.08-\e*\a-\offset,-0.02) node[midway, below] {$r_m$};
    
    \draw[red] plot[domain=-1:-0.725] ({-\a*cosh(\x)},{\b*sinh(\x)});
    \draw[red] plot[domain=0.725:1] ({-\a*cosh(\x)},{\b*sinh(\x)});
    \draw[->] plot[domain=-1.3:-1] ({-\a*cosh(\x)},{\b*sinh(\x)});
    \draw plot[domain=1:1.3] ({-\a*cosh(\x)},{\b*sinh(\x)});
    \draw[->] plot[domain=-0.725:0.725] ({-\a*cosh(\x)},{\b*sinh(\x)});
    
    \draw[dashed] plot[domain=-1.2:0] ({1.65*\x},{0.85*\x+0.22});
    \draw[dashed] plot[domain=-1.2:0] ({1.65*\x},{-0.85*\x-0.22});
    
    \def\xs1{-0.22/0.85}
    \def\xf1{-0.158}
    \draw[thick,->] (1.65*\xs1,0.85*\xs1+0.22) -- (1.65*\xf1,0.85*\xf1+0.22) node[above] {$\mathbf v^-_\infty$};
    \def\xs2{-0.22/0.85}
    \def\xf2{-0.36}
    \draw[thick,->] (1.65*\xs2,-0.85*\xs2-0.22) -- (1.65*\xf2,-0.85*\xf2-0.22) node[above] {$\mathbf v^+_\infty$};
    
    \draw [->,domain=30:150] plot ({-1.65*0.22/0.85+0.1*cos(\x)}, {0.1*sin(\x)}) node[right] {$2\delta$};

\end{tikzpicture}
\caption{The radial fly-by geometry. In the range $r\in[r_i, r_o]$, the spacecraft maintains an additional radial acceleration $\alpha$ and is thus not flying along a  hyperbola \label{fig:radialflyby}}
\end{center}
\end{figure}
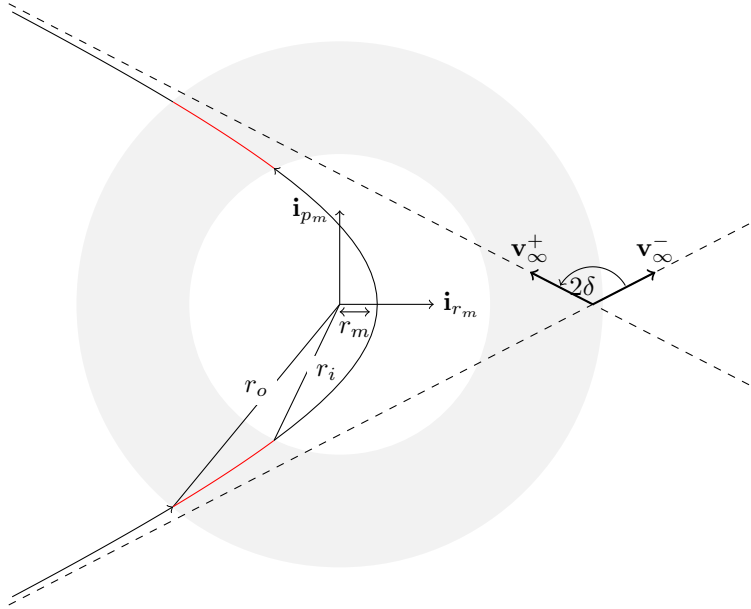

Planetary fly-bys are often modelled as Keplerian hyperbolas. In the preliminary mission design phases it is common to consider the effect of a fly-by as that of an instantaneous rotation of the relative velocity vector by the angle $\delta_K$ function of the closest passage distance $r_m$ and of the hyperbolic trajectory plane orientation. The outgoing relative velocity vector is then summed to the planet velocity vector to obtain the new spacecraft state in the interplanetary medium. The spacecraft is generally considered to be not thrusting during this phase. Here we study a simple powered fly-by manoeuvre where the spacecraft, along its planetocentric hyperbola, turns on its propulsion system when its distance $r$ from the planet is in the interval $[r_o, r_i]$ to produce a constant acceleration of magnitude $\alpha$. Since the resulting trajectory is perfectly symetric, also the effect of such a fly-by can be considered as a rotation of the relative velocity vector by an angle $\delta$. Using the analytical explicit solution in terms of the Weierstrass elliptic and related functions allow to derive simple equations to find $\delta$ as a function of $r_m, V_\infty, r_i, r_o$ and $\alpha$. In Figure \ref{fig:radialflyby} the basic geometry of a radial fly-by is shown. Note that the trajectory is not an hyperbola and is obtained by patching hyperbolic arcs (outside $[r_i, r_o]$) with the constant radial acceleration solution (in $[r_i, r_o]$).

\subsection{The unpowered flyby case $\alpha=0$}
First, as a benchmark, consider the case of a purely ballistic fly-by modelled as a Keplerian hyperbola. Assume as incoming conditions $v_{\infty}$, and write the specific energy at the incoming conditions at infinite:
$$
\mathcal E_K = \frac{v_\infty^2}{2}
$$
and at the closest distance:
$$
v_m^2 = \frac{2\mu}{r_m} + v_\infty^2
$$
where the subscript $K$ indicates that we are in the purely Keplerian case. We compute the angular momentum as $h_K = r_m v_m$:
$$
h_K = \sqrt{r_m(2\mu+v_\infty^2 r_m)}
$$
and the relative velocity rotation half angle as:
\begin{equation}
\sin \delta_K = \frac 1e
\label{eq:unpowered}
\end{equation}
where, accounting that $a(1-e^2) = \frac{h^2}{\mu}$, we have:
$$
e^2 = 1 + \frac{2h_K^2 \mathcal E_K}{\mu^2}
$$
so that from $r_m, v_\infty$ we can compute $\delta_K$.

\subsection{The powered flyby case}
We now study how the relative velocity rotation angle $\delta_K$ is modified when we assume a constant acceleration $\alpha$ acting on the spacecraft at $r \in [r_i, r_o]$ while keeping $r_m$ unchanged. From Figure \ref{fig:radialflyby} it can be seen how $2 \delta = 2 (\delta_1 + \delta_2 + \delta_3)$ is the sum of three contributions. The first and the last one, indicated with $2\delta_1$ and $2\delta_3$ are due to the hyperbolic motion outside the $[r_i, r_o]$ interval and can be computed using the Keplerian solution. The second contribution, indicated with $2 \delta_2$ is due to the radial accelerated motion and must thus be computed using Weierstrass functions. Let us preliminary determine the various motion invariants for the different arcs. Starting with the out-most hyperbolic arcs we have the specific energy:
$$
\mathcal E_{K_o} = \frac{v_\infty^2}{2}
$$
and by its conversation:
$$
v_o^2 =2 \left(\mathcal E_{K_o} + \frac \mu {r_o}\right)
$$
hence the specific energy along the following propelled arc will be:
$$
\mathcal E = \mathcal E_{K_o} - \alpha r_o
$$
and by its conversation:
$$
v^2_i = 2\left(\mathcal E_{K_o}+\frac\mu{r_i} +\alpha(r_i-r_o)\right)
$$
hence the energy along the inner hyperbolic arc will be:
$$
\mathcal E_{K_i} = \mathcal E_{K_o}+\alpha(r_i-r_o)
$$
and the velocity in the point of closest passage:
$$
v_m^2 = 2\left(\mathcal E_{K_o}+\frac\mu{r_m} +\alpha(r_i-r_o)\right)
$$
which allow to compute the angular momentum for all the arcs as $h(v_\infty, r_m, r_i, r_o) = r_m v_m$

\subsubsection{Computing $\delta_1$ and $\delta_3$}
To compute $\delta_1$ and $\delta_3$ we develop a generic expression for the angle $\delta$ between the velocities acquired along a Keplerian arc at two positions $r_1$ and $r_2$. Along a Keplerian hyperbola we have (see Battin\cite{battin} \S (3.6)):
$$
\frac{h_K}{\mu} \mathbf v =  - \sin \theta_K \mathbf{\hat i}_e + (e + \cos\theta_K) \mathbf{\hat i}_p
$$
where $\theta_K$ is the true anomaly in the Keplerian motion, i.e counted with respect to $ \mathbf{\hat i}_e$. We will drop the subscript $K$ for the true anomaly in the following expressions to avoid cluttering our notation. We thus have:
$$
\frac{p}{\mu}(\mathbf v_1 \cdot \mathbf v_2) = \frac{p}{\mu}v_1 v_2 \cos\delta = \sin\theta_1\sin\theta_2 + (e+\cos\theta_1)(e+\cos\theta_2)
$$
where the orbit polar $r = p / (1+e\cos\theta)$ can be used to compute:
$$
\begin{array}{ll}
\cos\theta = \frac 1e (\frac p{r} - 1) &
\sin\theta = -\frac 1e\sqrt{e^2-(\frac p{r} - 1)^2}
\end{array}
$$
where the sign minus is chosen on the sine as the hyperbolic arc considered is incoming. Applying the above formulae to compute $\delta_1$ and $\delta_3$ we get:
We finally have the the final expression for $\delta_1$:
$$
\frac{p}{\mu}v_\infty v_o \cos\delta_1 = \frac 1{e^2}\sqrt{(e^2-1)\left(e^2-\left(\frac p{r_o}-1\right)^2\right)}+\left(e-\frac 1e\right)\left(e + \frac 1e \left(\frac p{r_o} - 1\right)\right)
$$
and for $\delta_3$:
$$
\frac{p}{\mu}v_i v_m \cos\delta_3 = \left(e + \frac 1e \left(\frac p{r_i}-1\right)\right)(e+1)
$$
Note that in inverting the above expressions using $\arccos$ we have $\delta_1, \delta_3 \in [0, \pi]$ which is correct whenever no passage through the pericenter happens.

\subsubsection{Computing $\delta_2$}
While $r_i < r < r_o$ the propulsion system of the spacecraft turns on putting the spacecraft on a radially accelerated trajectory having $\mathcal E, h$ as motion invariants. We then may also compute $r_m$ and the lattice invariants $g_2$ and $g_3$ and hence all the relevant Weierstrass elliptic and related functions. 
We may then compute:
$$
\delta_2 = (\gamma_o - \theta_o) - (\gamma_i - \theta_i)
$$
where the true anomalies $\theta < 0$ are now referred to the non Keplerian arc and are computed by Eq.(\ref{eq:theta0}), while $\gamma \in [-\pi,0]$ is the flight path angle which can also be computed along the Keplerian arcs. We use:
$$
\tan\gamma = - \frac {r}{p}\sqrt{e^2-\left(\frac p{r} - 1\right)^2}
$$
The final effect of a radial fly-by will then be to rotate the relative velocity vector by an angle $2\delta = 2(\delta_1+\delta_2+\delta_3)$ by providing a cumulative $\Delta V_{lt} = 2 \alpha \cdot \mbox{tof}$ where $\mbox{tof} = t(\tau_o) - t(\tau_i)$ is the duration of the first propelled arc as computed applying Eq.(\ref{eq:radial_kepler}) in correspondence of the two radial anomalies at $r_i$ and $r_o$.

\subsubsection{A numerical example}
\begin{figure}[t!]
\centering
\includegraphics[width=0.95\columnwidth]{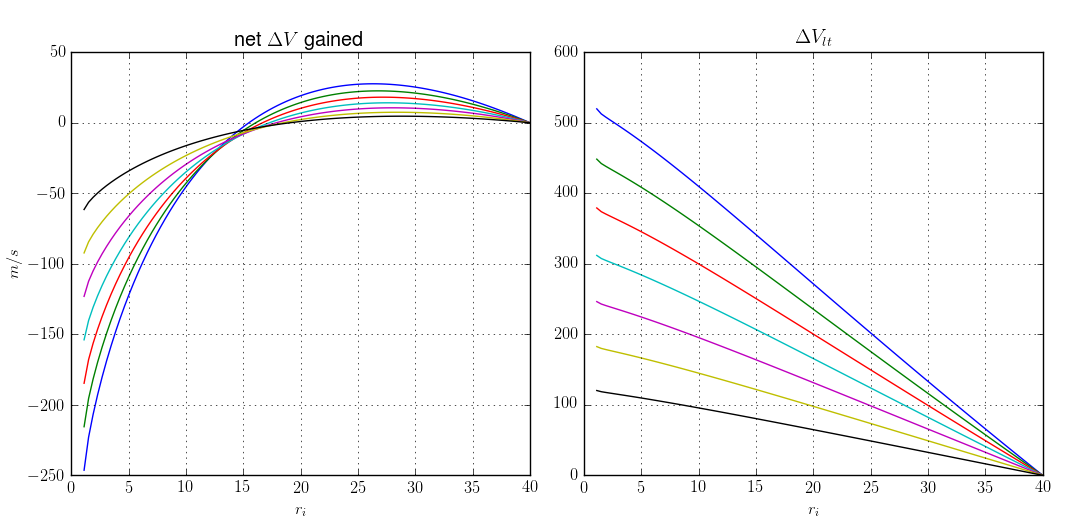}
\caption{Effect of a radial flyby at the Moon for $r_o=40 R_M$, $r_m = 1.1R_M$ and $m \alpha = [8, 7, 6, 5, 4, 3, 2]$ [N] with $m=2000$ [kg]. The net $\Delta V$ gained by the manoeuvre (left) is plotted aside the $\Delta V_{lt}$ necessary to perform the manoeuvre (right). \label{fig:moon_flyby}}
\end{figure}
Consider a Moon fly-by, where the spacecraft approaches the sphere of influence with a relative velocity $v_\infty = 1000$ [m/s] and performs a fly-by with closest approach distance $r_m = 1.1 R_M$, where $R_M = 4905$ [km] is the Moon radius. Under these conditions the effect of an unpowered fly-by, as computed from Eq.(\ref{eq:unpowered}), is that of rotating the relative velocity vector by an angle $2 \delta_K = 1.209$ [rad.]. We study the possibility to increase such an angle (the new value indicated by $2 \delta$) by performing a powered fly-by during which the spacecraft maintains a constant outward acceleration of magnitude $\alpha$ when its distance from the atracting body is $r\in[r_i, r_o]$. We consider to start the propelled phase at $r_o = 40 R_M$ and consider $r_i\in[1.1,40]R_M$. We compare the $\Delta V_{lt}$ used by the low-thrust propulsion system to the instantaneous $\Delta V_{K}$ that would be needed at the end of the outgoing asymptote to change the relative velocity direction by the same amount: $\Delta V_{K} = 2 v_\infty \sin{(\delta-\delta_K)}$. The angle $\delta$ as well as the $\Delta V_{lt}$ are computed using the formulae developed above. The use of the newly developed formulae enables to make this study very efficiently avoiding numerical propagation altogether. In Figure \ref{fig:moon_flyby} we plot the results in the selected case. We show the net gain of $\Delta V$ computed as the difference between $\Delta V_K$ and $\Delta V_{lt}$ as well as the value of $\Delta V_{lt}$. Note the area where a $\Delta V$ amplification effect is present. This is related to the decrease in spacecraft velocity which allows for the planet gravity to bend the asymptotes with greater efficiency. The analytical formulae derived express the deflection angle $\delta$ and the velocity increment $\Delta V_{lt}$ as an explicit function of $r_i, r_o, \alpha, v_\infty$. They are suitable to be used in a larger interplanetary trajectory optimization scheme, as well as in the preliminary assessment of some planetary encounter.

\section{Solution to the 2D Stark initial value problem}
Previous work \cite{stark} reported the explicit solution to the Stark problem in the full three dimensional case. That solution can be simplified if one restricts the problem to be purely two-dimensional. We present the derivation of the Stark problem solution in terms of Weierstrass functions specific for this simpler case. Note that an explicit solution using Jacobian elliptic functions is known for the 2D Stark case \cite{lantoine}. The reader is encouraged to compare the Weierstrass form of such a solution derived here to the solutions needed to cover all possible cases using Jacobi elliptic integrals. As recognized by Byrd\cite{byrd}, the Weierstrass approach has a clear advantage when the polynomial roots and their order are unknown, which is the case here as initial conditions will determine such roots hierarchy.

Consider the planar motion of a spacecraft subject to an inertially fixed acceleration of magnitude $\alpha > 0$ directed along the $x$ axis. The system specific energy, that is conserved, can be written as:
\begin{equation}
\mathcal E = \frac {v^2}2 - \frac \mu r - \alpha x
\label{eq:energystark}
\end{equation}
In a similar way as done in the full three dimensional case, we introduce the coordinates $\xi \in [-\infty, \infty]$, $\eta \in [-\infty, \infty]$ via the following transformations:
$$
\begin{array}{lr}
x = \frac{\xi^2 - \eta^2}2 & \dot x = \xi\dot\xi - \eta\dot\eta \\
y = \xi\eta & \dot y = \eta\dot\xi + \xi\dot\eta
\end{array}
$$
Note that with respect to the classical parabolic coordinates we do not restrict the domain of either $\eta$ or $\xi$ to the real axis. As a consequence, the transformation here used is not unique when reversed:
\begin{equation}
\begin{array}{lr}
\xi = \pm \sqrt{r + x} & \dot\xi = \pm \frac{\dot r + \dot x}{2\sqrt{r+x}} \\ 
\eta = \pm \sqrt{r - x}   & \dot\eta = \pm \frac{\dot r - \dot x}{2\sqrt{r-x}}
\end{array}
\label{eq:xietadef}
\end{equation}

where $r = \sqrt{x^2 + y^2} = \frac{\xi^2 + \eta^2}2$ and $\dot r = \frac{x\dot x + y\dot y}{r} = \dot\xi \xi + \dot\eta \eta$. This is not a problem here, rather an advantage, as we will only be interested in having unique values for $x$ and $y$. We may then choose any of the signs above when, for example, computing the initial conditions $\xi_0$ and $\eta_0$. 

Introduce now two further motion invariants $h_\xi$ and $h_\eta$\cite{stark}:
$$
\begin{array}{l}
h_\xi = -\alpha \frac {\xi^4}{2} - \mathcal E \xi^2 + 2 r^2\dot\xi^2 \\
h_\eta = \alpha \frac {\eta^4}{2} - \mathcal E \eta^2 + 2 r^2\dot\eta^2
\end{array}
$$
linked by the relation $h_\xi+h_\eta = 2\mu$. We may then write the fundamental differential equations that allow to solve the 2-D Stark problem:

\begin{equation}
\begin{array}{l}
\frac{d\xi}{d\tau} = \pm \sqrt{\alpha \xi^4 + 2\mathcal E\xi^2 + 2h_\xi} = \pm \sqrt{f_\xi}\\
\frac{d\eta}{d\tau} = \pm \sqrt{-\alpha \eta^4 + 2\mathcal E\eta^2 + 2h_\eta}  = \pm \sqrt{f_\eta}
\end{array}
\label{eq:differential_equations}
\end{equation}
where the pseudo-time $2r d\tau = dt$ is used.

\begin{figure}[t!]
\centering
\includegraphics[width=0.95\columnwidth]{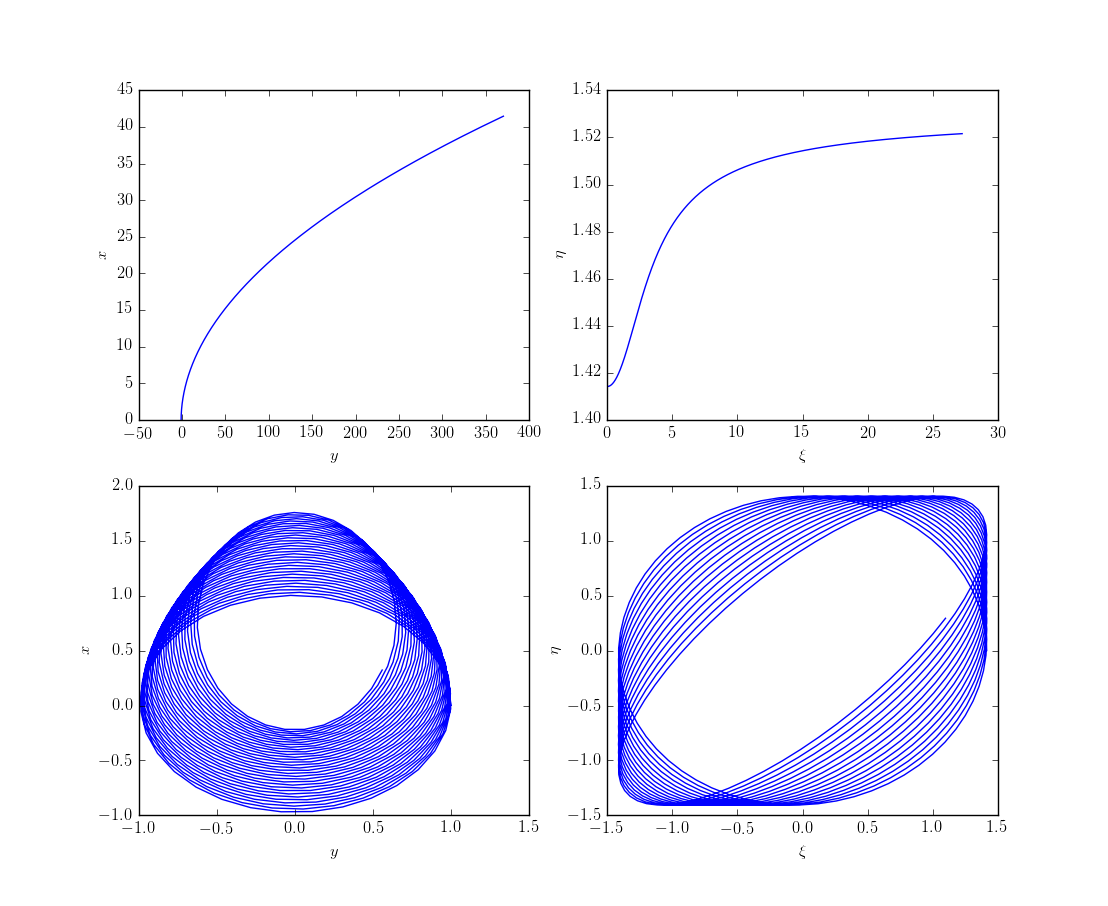}
\caption{Plot of the Stark 2D solution as obtained from Eq.(\ref{eq:xi}) and Eq.(\ref{eq:eta}) for an unbounded case (above) and a bounded case (below). Cartesian coordinates are shown together with one of the two possible choices for the parabolic ones. Initial conditions are $\mathbf r_0 = [-1.,1e-3]$, $\mathbf v_0=[1e-3,1.5]$, $\alpha = 0.1$ for the unbounded case, and $\mathbf r_0 = [1e-3,1]$, $\mathbf v_0=[1, 1e-3]$, $\alpha = 0.003$. Non dimensional units are used so that $\mu=1$.\label{fig:eta_xi}}
\end{figure}

\subsection{Roots of $f_\xi$}
It is straight forward to compute the roots of the polynomial $f_\xi$ applying the quadratic equation root formula on the fictitious variable $s = \xi^2$:
$$
s_{1,2} = \frac{-\mathcal E \pm \sqrt{\mathcal E^2-2\alpha h_\xi}}{\alpha}
$$
which translates immediately to the four solutions in $\xi$:
\begin{equation}
\tilde e^\xi_{1,2,3,4} = \pm \sqrt{\frac{-\mathcal E \pm \sqrt{\mathcal E^2-2\alpha h_\xi}}{\alpha}}
\label{eq:rootsxi}
\end{equation}
These roots can be, depending from the initial conditions, all complex, all real or two real and two pure imaginary.
\subsection{Roots of $f_\eta$}
Likewise, we compute the roots of the polynomial $f_\eta$ applying the quadratic equation root formula on the fictitious variable $s = \eta^2$:
$$
s_{1,2} = \frac{\mathcal E \pm \sqrt{\mathcal E^2+2\alpha h_\eta}}{\alpha}
$$
which translates immediately to the four solutions in $\eta$:
\begin{equation}
\tilde e^\eta_{1,2,3,4} = \pm \sqrt{\frac{\mathcal E \pm \sqrt{\mathcal E^2+2\alpha h_\eta}}{\alpha}}
\label{eq:rootseta}
\end{equation}

Note that $\lim_{\eta\to\pm\infty} f(\eta) = -\infty$ and $f(\eta_0) > 0$ since $\dot\eta_0$ must be real. As a consequence at least one of the above roots (and hence two) must be real.

\subsection{Solution for the $\xi$ coordinate}
Consider the first of Eqq.(\ref{eq:differential_equations}). Rewrite it as follow:
$$
\tau - \tau_{m,\xi} = \pm \int_{\xi_m}^\xi \frac {ds}{\sqrt{\alpha s^4 + 2\mathcal Es^2 + 2h_\xi}}
$$
where $\xi_m$ is any one of the four roots in Eq.(\ref{eq:rootsxi}). 
We recognize the expression has the form  Eq.(\ref{eq:weierstrass1}) and we thus 
introduce the lattice invariants, as defined in Eq.(\ref{eq:lattice_invariants}):
\begin{equation}
\begin{array}{l}
g_{2, \xi} = 2\alpha h_\xi + \frac{\mathcal E^2}{3}\\
g_{3, \xi} = \frac {\mathcal E}{3}\left( 2\alpha h_\xi - \frac {\mathcal E^2}9\right)
\end{array}
\label{eq:xiinv}
\end{equation}
and find the lattice roots as:
$$
e_{1,\xi} = - \frac {\mathcal E} 3, \quad
e_{2,\xi} = \frac {\mathcal E} 6 + \frac{\sqrt 2}{2}\sqrt{\alpha h_\xi}, \quad
e_{3, \xi} = \frac {\mathcal E} 6 - \frac{\sqrt 2}{2}\sqrt{\alpha h_\xi}
$$
Apply now Eq.(\ref{eq:weierstrass2}) to write:
\begin{equation}
\xi(\tau) = \xi_m + \frac{A_\xi}{\wp_\xi(\tau-\tau_{m,\xi})-B_\xi}
\label{eq:xi}
\end{equation}
where $A_\xi = 1 / 4 f'(\xi_m) = \xi_m(\alpha \xi_m^2 + \mathcal E)$, $B_\xi = 1 / 24 f''(\xi_m) = \frac{1}{2}(\alpha\xi_m^2 + \frac 13 \mathcal E)$ and $\wp_\xi(\tau_{m,\xi}) = B_\xi + \frac{A_\xi}{\xi_0 - \xi_m}$, $\wp_\xi'(\tau_{m,\xi}) = \frac{2 r_0 A_\xi \dot\xi_0 }{(\xi_0 - \xi_m)^2}$. We also have:
\begin{equation}
\dot\xi(\tau) = - \frac{(\xi(\tau) - \xi_m)^2}{2rA_\xi}\wp_\xi'(\tau - \tau_{m,\xi})
\label{eq:dxi}
\end{equation}
The expressions derived are valid regardless of the choice of the root $\xi_m$, but as explained previously, choosing a real root is desirable. The conditions to have at least one real root for the polynomial $f$, as easily verified from Eq.(\ref{eq:rootsxi}), can be written as:
\begin{equation}
\begin{array}{l}
a) \quad \mathcal E>0, \alpha h_\xi <0 \\
b) \quad \mathcal E<0, \alpha h_\xi<\frac{\mathcal E^2}{2}
\end{array}
\end{equation}
in both cases $\tilde e_1 = \sqrt{\frac{- \mathcal E + \sqrt{\mathcal E^2-2\alpha h_\eta}}{\alpha}}$ will result to be the expression for one of the real roots and we thus choose it in all cases. In essence, $h_\xi$ determines whether at least one real root exist. Noting that $\lim_{\alpha\to\infty} h_\xi = -\frac{\alpha}{2}y^2$ we can be certain that high level of thrust put us in this condition. Looking into the opposite direction, we note how low thrust levels also guarantee that, for $\mathcal E < 0$, a real root exists.

\subsection{Solution for the $\eta$ coordinate}
Consider the second of Eq.(\ref{eq:differential_equations}). Rewrite it as follows:
$$
\tau - \tau_{m,\eta} = \pm \int_{\eta_m}^\eta \frac {ds}{\sqrt{-\alpha s^4 + 2\mathcal Es^2 + 2h_\eta}}
$$
where $\eta_m$ is any one of the four roots in Eq.(\ref{eq:rootsxi}). 
We recognize the expression has the form  Eq.(\ref{eq:weierstrass1}) and we thus 
introduce the lattice invariants, as defined in Eq.(\ref{eq:lattice_invariants}):
\begin{equation}
\begin{array}{l}
g_{2, \eta} = -2\alpha h_\eta + \frac{\mathcal E^2}{3}\\
g_{3, \eta} = \frac {\mathcal E}{3}\left( -2\alpha h_\eta - \frac {\mathcal E^2}9\right)
\end{array}
\label{eq:etainv}
\end{equation}
and find the lattice roots as:
$$
e_{1,\eta} = - \frac {\mathcal E} 3, \quad
e_{2,\eta} = \frac {\mathcal E} 6 + \frac{\sqrt 2}{2}\sqrt{-\alpha h_\eta}, \quad
e_{3, \eta} = \frac {\mathcal E} 6 - \frac{\sqrt 2}{2}\sqrt{-\alpha h_\eta}
$$
Apply now Eq.(\ref{eq:weierstrass2}) to write:
\begin{equation}
\eta(\tau) = \eta_m + \frac{A_\eta}{\wp_\eta(\tau-\tau_{m,\eta})-B_\eta}
\label{eq:eta}
\end{equation}
where $A_\eta = 1 / 4 f'(\eta_m) = \eta_m(-\alpha \eta_m^2 + \mathcal E)$, $B_\eta = 1 / 24 f''(\eta_m) = \frac{1}{2}(-\alpha\eta_m^2 + \frac 13 \mathcal E)$ and $\wp_\eta(\tau_{m,\eta}) = B_\eta + \frac{A_\eta}{\eta_0 - \eta_m}$, $\wp_\eta'(\tau_{m,\eta}) = \frac{2 r_0 A_\eta \dot\eta_0 }{(\eta_0 - \eta_m)^2}$. 
We also have:
\begin{equation}
\dot\eta(\tau) = - \frac{(\eta(\tau) - \eta_m)^2}{2rA_\eta}\wp_\eta'(\tau - \tau_{m,\eta})
\label{eq:deta}
\end{equation}
It is possible to show that $\tilde e^\eta_1$ is always real and we will thus choose it as our $\eta_m$

\subsection{The time equation}
The pseudo time $\tau$ is defined via the differential equation $dt = (\xi^2 + \eta^2) d\tau$ which we may now write explicitly using Eq.(\ref{eq:xi}) and Eq.(\ref{eq:eta}):
$$
\frac{dt}{d\tau} = \xi_m + \eta_m + \frac{A_\xi^2}{[\wp(\tau-\tau_{m,\xi})-B_\xi]^2}  + \frac{A_\eta^2}{[\wp(\tau-\tau_{m,\eta})-B_\eta]^2} + \frac{2A_\xi}{\wp(\tau-\tau_{m,\xi})-B_\xi}  + \frac{2A_\eta}{\wp(\tau-\tau_{m,\eta})-B_\eta}
$$
In order to integrate the above equation, we employ two formulae from Tannery \& Molk \cite{jules_tannery_elements_1893}  [Chapter CXII] (see also Gradshte\u{\i}n \& Ryzhik\cite{gradshtein_table_2007} [\S 5.141]):
\begin{align}
\int\frac{du}{\wp\left(u\right) - \wp\left(v\right)} & = \frac{1}{\wp^\prime\left(v\right)}\left[\ln\frac{\sigma\left(u - v\right)}{\sigma\left(u + v\right)}+2u\zeta\left(v\right)\right],\label{eq:molk_00}\\
\int\frac{du}{\left[\wp\left(u\right) - \wp\left(v\right)\right]^2} & =  - \frac{1}{\wp^{\prime 2}\left(v\right)} \left[
\zeta\left(u - v\right)+\zeta\left(u + v\right)
\vphantom{\int\frac{du}{\wp\left(u\right) - \wp\left(v\right)}}
\right.\notag\\
&\quad \left. + 2u\wp\left(v\right)+\wp^{\prime\prime}\left(v\right) \int\frac{du}{\wp\left(u\right) - \wp\left(v\right)}
\right].\label{eq:molk_01}
\end{align}
where it is assumed that $\wp'(v) \ne 0$, that is $v$ is not a root of the Weierstrass polynomial $g(s) = 4s^3-g_2s-g_3$.
We introduce the shorthand notation:
\begin{align}
\mathcal{J}_1\left(u,v\right) & = \int\frac{du}{\wp\left(u\right) - \wp\left(v\right)},\\
\mathcal{J}_2\left(u,v\right) & = \int\frac{du}{\left[\wp\left(u\right) - \wp\left(v\right)\right]^2}
\end{align}
(with the understanding that we will add a $\xi$ or $\eta$ subscript depending on the subscript of the Weierstrassian functions appearing in the integrals). We may then derive the following time equation:
\begin{align}
t(\tau) = &(\xi_m + \eta_m) \tau + \notag\\
&A^2_\xi \left(\mathcal{J}_{2,\xi}\left(\tau-\tau_{m,\xi},b_\xi\right)+ \mathcal{J}_{2,\xi}\left(\tau_{m,\xi},b_\xi\right) \right) + \notag\\
&A^2_\eta \left(\mathcal{J}_{2,\eta}\left(\tau-\tau_{m,\eta},b_\eta\right)+ \mathcal{J}_{2,\eta}\left(\tau_{m,\eta},b_\eta\right) \right) + \notag\\
&2A_\xi \left(\mathcal{J}_{1,\xi}\left(\tau-\tau_{m,\xi},b_\xi\right)+ \mathcal{J}_{1,\xi}\left(\tau_{m,\xi},b_\xi\right) \right) + \notag\\
&2A_\eta \left(\mathcal{J}_{1,\eta}\left(\tau-\tau_{m,\eta},b_\eta\right)+ \mathcal{J}_{1,\eta}\left(\tau_{m,\eta},b_\eta\right) \right)
\end{align}
where we have defined $b_\xi = \wp_\xi^{-1}(B_\xi)$, $b_\eta = \wp_\eta^{-1}(B_\eta)$ (note that we can take any of the values returned by the inverse $\wp$ as $\mathcal J_1(u,v) = \mathcal J_1(u,-v)$, $\mathcal J_2(u,v) = \mathcal J_2(u,-v)$).

\section{Stark fly-by}
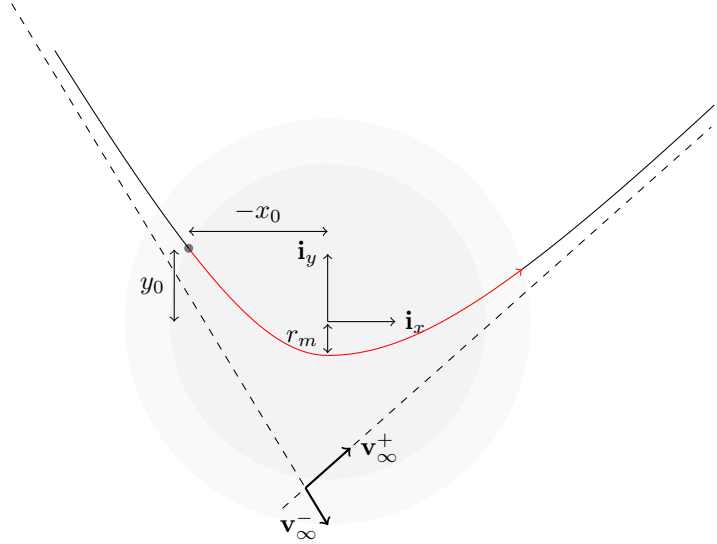
\begin{figure}[t!]
\begin{center}
\begin{tikzpicture}[scale=3]
    \pgfmathsetmacro{\e}{1.15}   
    \pgfmathsetmacro{\a}{1}
    \pgfmathsetmacro{\b}{(\a*sqrt((\e)^2-1)} 
    
    \pgfmathsetmacro{\ee}{1.4}   
    \pgfmathsetmacro{\aa}{(\a*\e) / \ee}
    \pgfmathsetmacro{\bb}{(\aa*sqrt((\ee)^2-1)}

    \path [draw=none,fill=gray, fill opacity = 0.05] (0,0) circle (0.7);
    \path [draw=none,fill=gray, fill opacity = 0.05] (0,0) circle (0.9);  
    
    \draw[->] (0,0) -- (0.3,0) node[right] {$\mathbf i_{x}$};
    \draw[->] (0,0) -- (0,0.3) node[left] {$\mathbf i_{y}$};
    
    
        
    
    \draw[red] plot[domain=-0.94:0] ({\b*sinh(\x)}, {\a*cosh(\x) - \e*\a});
    \draw[red, ->] plot[domain=0:0.93] ({\bb*sinh(\x)}, {\aa*cosh(\x) + (\a-\aa) - \e*\a});
    \draw[] plot[domain=-1.5:-0.94] ({\b*sinh(\x)}, {\a*cosh(\x) - \e*\a});
    \draw[] plot[domain=0.93:1.5] ({\bb*sinh(\x)}, {\aa*cosh(\x) + (\a-\aa) - \e*\a});
    \path [draw=none,fill=black, fill opacity = 0.5] ({\b*sinh(-0.94)}, {\a*cosh(-0.94) - \e*\a}) circle (0.02);
   
    \pgfmathsetmacro{\A}{-1.65}
    \pgfmathsetmacro{\B}{-0.9}
    \draw[dashed] plot[domain=-1.4:0] ({\x},{\A*\x+\B});
    \pgfmathsetmacro{\C}{0.89}
    \pgfmathsetmacro{\D}{-0.65}
    \draw[dashed] plot[domain=-0.2:1.7] ({\x},{\C*\x+\D});
    
    \pgfmathsetmacro{\xv}{ - (\B-\D) / (\A-\C)}
    \pgfmathsetmacro{\yv}{\A*\xv+\B}

    \draw[thick,->] (\xv,\yv) -- (\xv+0.1,\yv+\A*0.1) node[left] {$\mathbf v^-_\infty$};
    \draw[thick,->] (\xv,\yv) -- (\xv+0.2,\yv+\C*0.2) node[right] {$\mathbf v^+_\infty$};

    \draw[<->] (-0.615,0.4) -- (0,0.4) node[midway, above] {$-x_0$} ;
    \draw[<->] (-0.68,0.32) -- (-0.68,0) node[midway, left] {$y_0$} ;
    \draw[<->] (0,-0.01) -- (0,-0.14) node[midway, left] {$r_m$} ;

\end{tikzpicture}
\caption{The Stark fly-by geometry. \label{fig:starkflyby}}
\end{center}
\end{figure}
Consider a second case of powered flyby (which we will refer to as to a Stark fly-by) where the spacecraft, incoming along a purely ballistic trajectory (hyperbola) ignites his propulsion system at $\mathbf r_0=[x_0, y_0]$ to keep a constant acceleration $\alpha$ inertially fixed along the direction $\mathbf i_x$. The spacecraft propelled arc targets a closest planetary distance of $r_m$ and the spacecraft cuts off its propulsion system after a time $T$. It is known that a single velocity increment delivered at the pericenter along the direction of the velocity vector is an efficient way to increase the spacecraft velocity as the final velocity increment obtained at the end of the outbound hyperbola arc results to be much higher than the one delivered. To leverage this effect, often  known as the Oberth effect, we require the fixed constant acceleration direction to be aligned with the spacecraft velocity at the closest passage $r_m$. Such an alignment will be lost as the constant acceleration direction is kept fixed while the spacecraft velocity will be bended by the planet gravity. The fly-by trajectory geometry for this case is shown in Figure \ref{fig:starkflyby}. The use of this manoeuvre as part of an interplanetary trajectory can be studied once an efficient and simple procedure to design it is laid down. Assume $v_\infty$ and $r_m$ as known, and consider the problem of designing a Stark fly-by such that the velocity at the pericenter is $v_m$. Choosing the variable $v_m$ to parametrize the Stark fly-by allows to compute the lattice invariants only once and thus computations that make use of Weierstrass functions are extremely efficient as discussed in Section \S \ref{sec:computational efficiency}.

Start computing the velocity at the pericenter $v_{mK}$ along a Keplerian hyperbola defined by the same entry condition $v_\infty$ and pericenter distance $r_m$:
$$
\frac{v_{mK}^2}{2} = \frac{v_\infty^2}{2} + \frac{\mu}{r_m}
$$
The energy along the propelled arc is obtained from Eq.(\ref{eq:energystark}) applied at the $y$ axis crossing (i.e. $x=0$):
\begin{equation}
\mathcal E = \frac{v_m^2}{2} - \frac{\mu}{r_m}
\label{eq:energystarkx0}
\end{equation}
which allows to compute $x_0$ from:
$$
\frac{v_\infty^2}{2}-\alpha x_0 = \mathcal E
$$
Note that $x_0$ will be negative if $v_m > v_{mK}$ as the spacecraft will need to accelerate, while 
$x_0$ will be positive if $v_m < v_{mK}$ as the spacecraft will need to decelerate. Compute now the motion invariants $h_\xi$ and $h_\eta$ at $x=0$. In this point $\xi=-\eta=\sqrt{r_m}$ and the motion invariants $h_\xi$ and $h_\eta$ can be written as:
$$
\begin{array}{ll}
h_\xi = -\alpha \frac {r_m^2}{2} - \mathcal E r_m + 2 r_m^2\dot\xi_m^2 \\
h_\eta = \alpha \frac {r_m^2}{2} - \mathcal E r_m + 2 r_m^2\dot\eta_m^2
\end{array}
$$
Consider now the $x=0$ point as the pericenter of the propelled arc (in order to maximize the Oberth effect). From Eq.(\ref{eq:xietadef}) it is easy to derive $\dot \xi^2 = \frac{v_m^2}{4r_m}$, $\dot \eta^2 = \frac{v_m^2}{4r_m}$ since $\dot x = v_m$ and $\dot r= 0$ in this case. Hence we get the following expressions for the motion invariants :
\begin{equation}
\begin{array}{ll}
h_\xi = -\alpha\frac{r_m^2}{2} + \mu \\
h_\eta = \alpha\frac{r_m^2}{2} + \mu
\end{array}
\label{eq:motinvfb}
\end{equation}

\begin{figure}[t!]
\centering
\includegraphics[width=0.7\columnwidth]{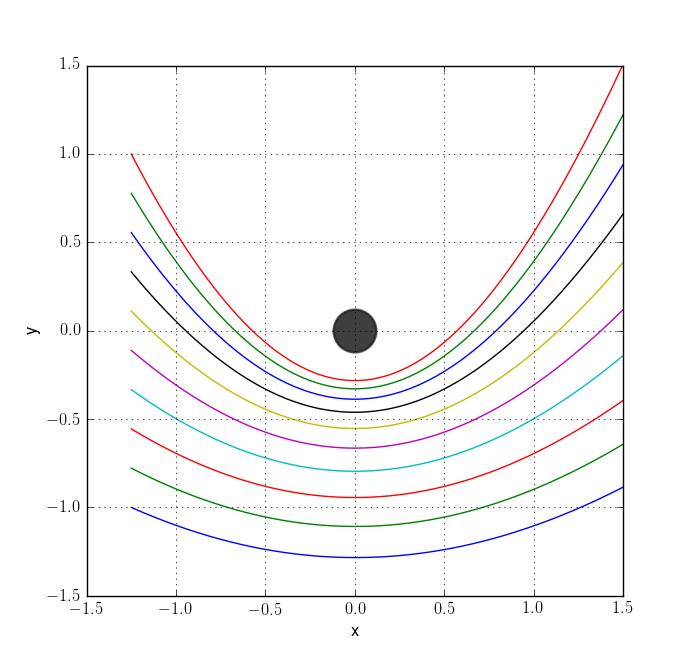}
\caption{Stark fly-bys with motion invariants $\mathcal E = 5.125\cdot 10^{-2}$, $h_\xi=0.9975$, $h_\eta=1.0025$ and $v_\infty = 0.3$ ($v_{m} = 1.45 > v_{mK}$). The $x_0$ coordinate at the beginning of the manoeuvre (left) is determined, while the $y_0$ coordinate is free to vary. \label{fig:stark_fb1}}
\end{figure}
We may now compute the lattice invariants $g_{2,\xi}$, $g_{3,\xi}$ from Eq.(\ref{eq:xiinv}) and $g_{2,\eta}$, $g_{3,\eta}$ from Eq.(\ref{eq:etainv}). The flyby geometry is, though, still not fully determined as there are infinitely many propelled arcs, parametrized by the other initial condition $y_0$, having the computed motion invariants and satisfying the entry condition on $v_\infty$. Figure \ref{fig:stark_fb1} visualized the different arcs parametrized by different $y_0$. The initial value $y_0$ can be selected by forcing at $x=0$ the distance from the origin to be equal to the requested $r_m$. Since at $x=0$, $\xi^2 = r$ this requires solving the equation $\xi^2(\tau^*) = r_m$. The value $\tau^*$ is thus found solving the equation $\xi(\tau^*)=\eta(\tau^*) \to x=0$. This is done using Eq.(\ref{eq:xi}) and Eq.(\ref{eq:eta}) to compute $\xi(\tau^*)$ and $\eta(\tau^*)$ and a simple Newton iteration to get $\tau^*$ (as initial guess $\tau^*=0$ reveals to be good in most cases). 

\subsection{Pseudo-algorithm to design a Stark fly-by}
Assume to know the values $\mu, \alpha, v_\infty, r_m, v_m$. Compute $\mathcal E$ from Eq.(\ref{eq:energystarkx0}) and $h_\xi$, $h_\eta$ from Eq.(\ref{eq:motinvfb}). Compute the lattice invariants from Eq.(\ref{eq:xiinv}) and Eq.(\ref{eq:etainv}). Assume a value for $y_0$. At the start of the propelled arc we have:
$$
\begin{array}{l}
x_0 = \frac 1{\alpha}\left( \frac{v_\infty^2}{2} - \mathcal E\right) \\
r_0 = \sqrt{x_0^2 + y_0^2}
\end{array}
$$
Compute the parabolic coordinates at the starting point from:
$$
\begin{array}{ll}
\xi_0 = \sqrt{r_0+x_0} & \dot\xi_0 = - \frac{\sign{(y_0)}}{2r_0} \sqrt{\alpha \xi_0^4 + 2\mathcal E\xi_0^2 + 2h_\xi}\\
\eta_0 = \sign{(y_0)} \sqrt{r_0-x_0} & \dot\eta_0 = - \frac{\sign{(y_0)}}{2r_0} \sqrt{-\alpha \eta_0^4 + 2\mathcal E\eta_0^2 + 2h_\eta}
\end{array}
$$
where we have used Eq.(\ref{eq:differential_equations}) and we have chosen the signs so that the concavity of the represented shape is positive. Find $\tau^*$ using Newton iterations to solve the equation:
$$
\xi(\tau^*)=\eta(\tau^*)
$$
where $\xi$ and $\eta$ are given by Eq.(\ref{eq:xi}) and Eq.(\ref{eq:eta}) respectively. Note that $\tau_{m,\xi}$ and $\tau_{m,\eta}$ need to be computed for each assumed $y_0 \to \eta_0, \xi_0$. Iterate using a Newton method on the assumed $y_0$ so that the following relation is satisfied:
$$
\xi^2(\tau^*) = r_m
$$
Having determined $x_0$ and $y_0$ all the remaining relevant quantities can be easily found solving an initial value problem.

\subsection{A numerical example}
\begin{figure}[t!]
\centering
\includegraphics[width=\columnwidth]{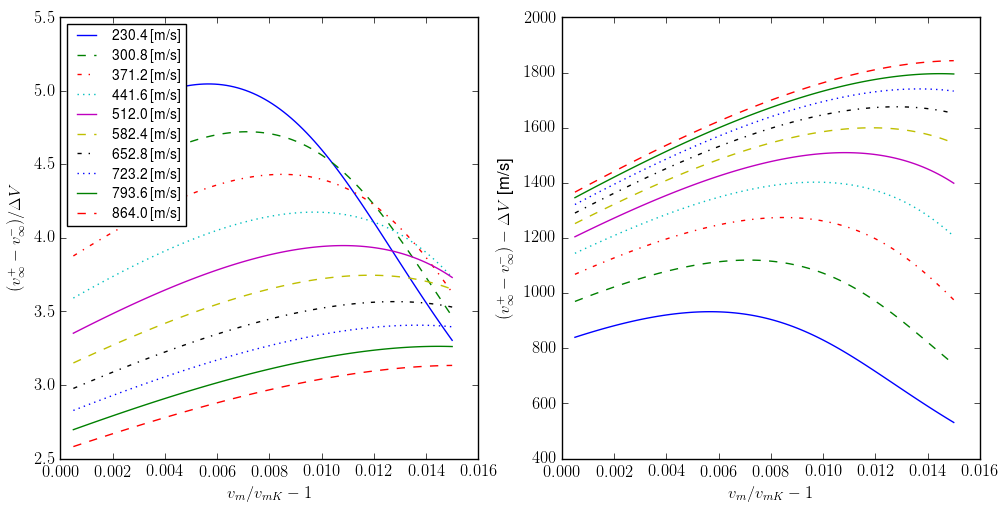}
\caption{The amplification factor during a Stark fly-by at Jupiter at different $\Delta V$ levels (left). The net gain is also shown (right). A Rosetta type of spacecraft is assumed with $m = 3000$ kg and $T=10$ [N] resulting in $\alpha=0.00\overline 3$. Incoming conditions of $v_\infty = 3000$ [m/s] are assumed and a closest passage radius of $r_m = 10 R_J$.\label{fig:starkfbperf}}
\end{figure}

A numerical example where the formulas derived above reveal their use is shown. We look into a Jupiter fly-by, where the spacecraft approaches Jupiter sphere of influence with a relative velocity $v_\infty = 3000$ [m/s] and performs a Stark fly-by with closest approach distance $r_m = 10 R_J$, where $R_J = 71492$ [km] is the Jupiter radius. The spacecraft, inspired by the Rosetta spacecraft, has a starting mass $m = 3000$ kg and a maximum thrust capability of $T_{max}=10$ [N] resulting in the possibility to apply a constant $\alpha=0.00\overline 3$. We assume a $\Delta V_{lt} = \alpha \Delta T$ available and we compute the velocity increment $v_\infty^+ - v_\infty^-$ at infinite  resulting from a Stark fly-by. The amplification factor, defined as $(v_\infty^+ - v_\infty^-) / \Delta V_{lt}$ as well as the net gain defined as $(v_\infty^+ - v_\infty^-) - \Delta V_{lt}$ is shown in Figure \ref{fig:starkfbperf} for different $\Delta V_{lt}$ and assumed $v_m$ magnitudes. 

\section{The Euler's (two-fixed centers) problem and Vinti's problem}
The same solution method reported above for the constant radial acceleration case and detailed above for the 2D Stark problem, can be used to solve the two-fixed centers, or Euler, problem\cite{euler}. While the formulae for the time equation and for the out-of-plane movement turns out to be more complicated, the underlying \lq\lq machinery\rq\rq\ is still based on the same few relations. In the present contribution we will not go in the details of those formulae, it is here sufficient to know that those formulae exist expressing the explicit solution to the Euler problem in its own pseudo-time (as before the solution with respect to time requires solving numerically a time equation expressed, also, in terms of the Weierstrass functions). We here will only remind that, following Aksenov et al.\cite{aksenov}, the solution of the two-fixed centers is linked to Vinti's problem\cite{vinti}, so that as we will briefly outline here, having an explicit solution to Euler's problem entails having an explicit analytical solution for the motion of an artificial satellite around an oblate primary, including the $J_2$ and $J_3$ terms.

Indicating with $\mu_1$ and $\mu_2$ the gravity parameters of the two bodies, we recognize that the system energy is:
$$
\mathcal E = \mathcal T + \mathcal V = \frac{\dot{\mathbf r}^2}2 - \frac{\mu_1}{r_1} - \frac{\mu_2}{r_2} 
$$
where, with respect to an inertial system having its origin in the center of mass of the system made by the two attracting bodies alone, we have
$$
\begin{array}{l}
r_1 = \sqrt{x^2+y^2+(z-a_1)^2} \\
r_2 = \sqrt{x^2+y^2+(z-a_2)^2} \\
\end{array}
$$
having indicated with $a_1$ and $a_2$ the distances of the two masses from the system origin. We may then expand the inverse distances appearing in the gravitational potential is series of Legendre polynomials obtaining the expansions:
$$
\begin{array}{l}
\frac 1{r_1} = \frac 1r \sum_{\ell=0}^\infty \left(\frac{a_1}{r}\right)^\ell P_\ell\left(\frac zr\right)\\
\frac 1{r_2} = \frac 1r \sum_{\ell=0}^\infty \left(\frac{a_2}{r}\right)^\ell P_\ell\left(\frac zr\right)\\
\end{array}
$$
and thus proving that the gravitational potential of the two-fixed centers problem has the form:
\begin{equation}
\mathcal V = - \frac{\mu_1+\mu_2}{r}\left(1 + \sum_{\ell=1}^\infty \frac{\gamma_\ell}{r^\ell} P_\ell\left(\frac zr\right)\right)
\end{equation}
where:
$$
\gamma_\ell = \frac{\mu_1 a_1^\ell + \mu_2 a_2^\ell}{\mu_1+\mu_2}
$$
Since our reference system has its origin at the center of mass of the two bodies, we have $\gamma_1 = 0$ and thus we can take the sum from $\ell=2$. Lets compare the expression above to the gravitational potential of an axisymmetric spheroid:
$$
\mathcal V_{sph} = - \frac{\mu}{r}\left(1 - \sum_{\ell=2}^\infty J_\ell \left(\frac{R_e}{r}\right)^\ell P_\ell\left(\frac zr\right)\right)
$$
Assuming $\mu_1$, $\mu_2$, $a_1$ and $a_2$ as complex numbers of the form:
$$
\begin{array}{cc}
 \mu_1 = \frac \mu 2 (1+\imath\sigma)& a_1 = c (\sigma + \imath) \\
 \mu_2 = \frac \mu 2 (1-\imath\sigma)& a_2 = c (\sigma - \imath)
\end{array}
$$
we are guaranteed that the potential $\mathcal V$ will be real\cite{aksenov}. Writing $J_2 R_e^2 = \gamma_2$ and $J_3 R_e^3 = \gamma_3$ the following conditions are found:
$$
\begin{array}{c}
c^2 (1+\sigma^2) = - J_2 \\
2\sigma c^3  (1+\sigma^2) = -J_3
\end{array}
$$
which allow the first terms of $\mathcal V$ to be equal to the first terms of $\mathcal V_{sph}$. The following term will then be $\gamma_4 = c^4(1+\sigma^2)(1-3\sigma^2)$ and cannot be made equal to $J_4 R_e^4$.

The above developments summarize the original idea from Aksenov et al.\cite{aksenov} who, though, did not develop the explicit solution further stopping at the derivation of the quadratures. The use of the Weierstrassian formalism here introduced allows to find such expressions (as shown in the classic case\cite{euler}) and thus to have a fully analytical solution to the problem of an artificial satellite motion around an oblate primary precise up to the $J_3$ term.

\section{Conclusion}
Weierstrass elliptic and related functions express via an elegant formalism solutions to fundamental problems in astrodynamics. These include the constant radial acceleration problem, the Stark problem and the Euler, two-fixed center problem. In all cases the same solution procedure can be applied, resulting in explicit solutions with respect to a Sundmann transformed pseudo-time (or anomaly when possible). The resulting new approach proves to be useful to describe powered fly-bys and the motion of an artificial satellite subject to the gravitational field of an oblate primary including $J_2$ and $J_3$ harmonics (Vinti's problem).

\bibliographystyle{aiaa}
\bibliography{weierstrass}

\end{document}